\documentclass{PoS}

\usepackage{lineno}
\usepackage{breakurl}
\usepackage[utf8]{inputenc}
 \usepackage{url}

\let\OLDthebibliography\thebibliography
\renewcommand\thebibliography[1]{
  \OLDthebibliography{#1}
  \setlength{\parskip}{0pt}
  \setlength{\itemsep}{0pt plus 0.3ex}
}

\title{Searching for VHE gamma-ray emission associated with IceCube astrophysical neutrinos using FACT, H.E.S.S., MAGIC, and VERITAS}

\ShortTitle{IACT follow-up of IceCube neutrino events}

\author{\speaker{M.~Santander} on behalf of the VERITAS Collaboration\thanks{http://veritas.sao.arizona.edu}}
\author{D. Dorner on behalf of the FACT Collaboration\thanks{http://www.fact-project.org/}}
\author{J. Dumm on behalf of the IceCube Collaboration\thanks{http://icecube.wisc.edu/collaboration/authors/icrc17\_icecube}}
\author{K. Satalecka on behalf of the MAGIC Collaboration\thanks{https://magic.mpp.mpg.de/}}
\author{F. Sch\"ussler on behalf of the H.E.S.S. Collaboration\thanks{https://www.mpi-hd.mpg.de/hfm/HESS/}}

\abstract{The detection of an astrophysical flux of high-energy neutrinos by IceCube is a major step forward in the search for the origin of cosmic rays, as this emission is expected to originate in hadronic interactions taking place in or near cosmic-ray accelerators. No neutrino point sources, or a significant correlation with known astrophysical objects, have been identified in the IceCube data so far that could reveal the location of the neutrino emission sites. The hadronic interactions responsible for the neutrino emission should also lead to the production of high-energy gamma rays from neutral pion decays. The search for neutrino sources can therefore be performed by studying the spatial and temporal correlations between neutrino events and very-high-energy (VHE, E > 100 GeV) gamma rays. We report on the search for VHE gamma-ray emission at the reconstructed position of muon neutrino events detected by IceCube using the FACT, H.E.S.S., MAGIC, and VERITAS imaging atmospheric Cherenkov telescopes (IACTs). No significant steady gamma-ray counterparts have been identified for the neutrino events observed so far. 
Finally, we outline recent programs to perform prompt IACT observations of realtime IceCube neutrino event positions.
}

\FullConference{35th International Cosmic Ray Conference - ICRC2017 \\
		10-20 July, 2017\\
		Bexco, Busan, Korea}

\begin{document}

\section{Introduction}

The origin of the astrophysical neutrino flux discovered by the IceCube~\cite{HESE1} observatory remains unknown. The astrophysical flux is significant at energies between $\sim20$ TeV and a few PeV and its energy spectrum is consistent with a $E^{-\Gamma}$ power law, with IceCube analyses reporting spectral indices $\Gamma$ in the 2.1--2.5 range~\cite{Leif}. While no neutrino point sources have been identified so far, the apparent isotropy of the astrophysical events and the lack of a significant correlation with the Galactic Plane seems to favor a dominant extragalactic component.

The search for the sources of astrophysical neutrinos can be performed using a multi-messenger approach by combining neutrino and gamma-ray observations. As the neutrinos are expected to be produced in the decay of pions originating in cosmic-ray interactions, the same process should give rise to a flux of hadronic gamma rays which could be detected at Earth if they are not attenuated by interactions within the source or during propagation. Over the last four years, the imaging atmospheric Cherenkov telescopes (IACTs) FACT\footnote{FACT is a proof-of-principle project for silicon based photosensors (SiPM) and has a higher threshold due to its smaller mirror area.}, H.E.S.S., MAGIC and VERITAS, sensitive to gamma rays in the very-high-energy band (VHE, E $>$ 100 GeV), have been utilized to search for gamma-ray emission associated with high-energy neutrino events detected by IceCube which are potentially astrophysical in origin. In this work, we present a status update on the search for neutrino event counterparts in the VHE gamma-ray band, presenting preliminary results from this program and discussing the constraints that these results could set on the properties of the neutrino sources. 

\section{Detectors and datasets}

The search for VHE gamma-ray emission is performed at the positions pointed back to by muon-track events detected by IceCube which, given their high energy, are more likely to be astrophysical than to be to be produced in an atmospheric cosmic-ray shower. IceCube is a cubic-kilometer neutrino detector installed in the ice at the geographic South Pole~\cite{IC3} between depths of 1450\,m and 2450\,m. Detector construction started in 2005 and finished in 2010. Neutrino reconstruction relies on the optical detection of Cherenkov radiation emitted by secondary particles produced in neutrino interactions in the surrounding ice or the nearby bedrock. Track events are used as their typical  $<1^{\circ}$ error circle can be covered using the $3.5^{\circ}-5^{\circ}$ field-of-view of current generation IACTs.  In this work, we focus on follow-up observations of neutrino positions selected from three different IceCube analyses: from a four-year search for high-energy events with their neutrino interaction vertices contained in the IceCube volume (HESE, \cite{HESE3}), a two-year search for high-energy $\nu_{\mu}$ events from the Northern hemisphere~(HEMU2, \cite{Chris}), and from an update on this last analysis using six years of data (HEMU6, \cite{Leif}). As the HEMU6 analysis includes the period covered by HEMU2, some overlap exists between the events in the two data sets. The overlap is however not complete, given the difference in the event selection criteria applied in both analyses. 

\begin{figure}[!th]
\centering
\includegraphics[width=0.95\textwidth]{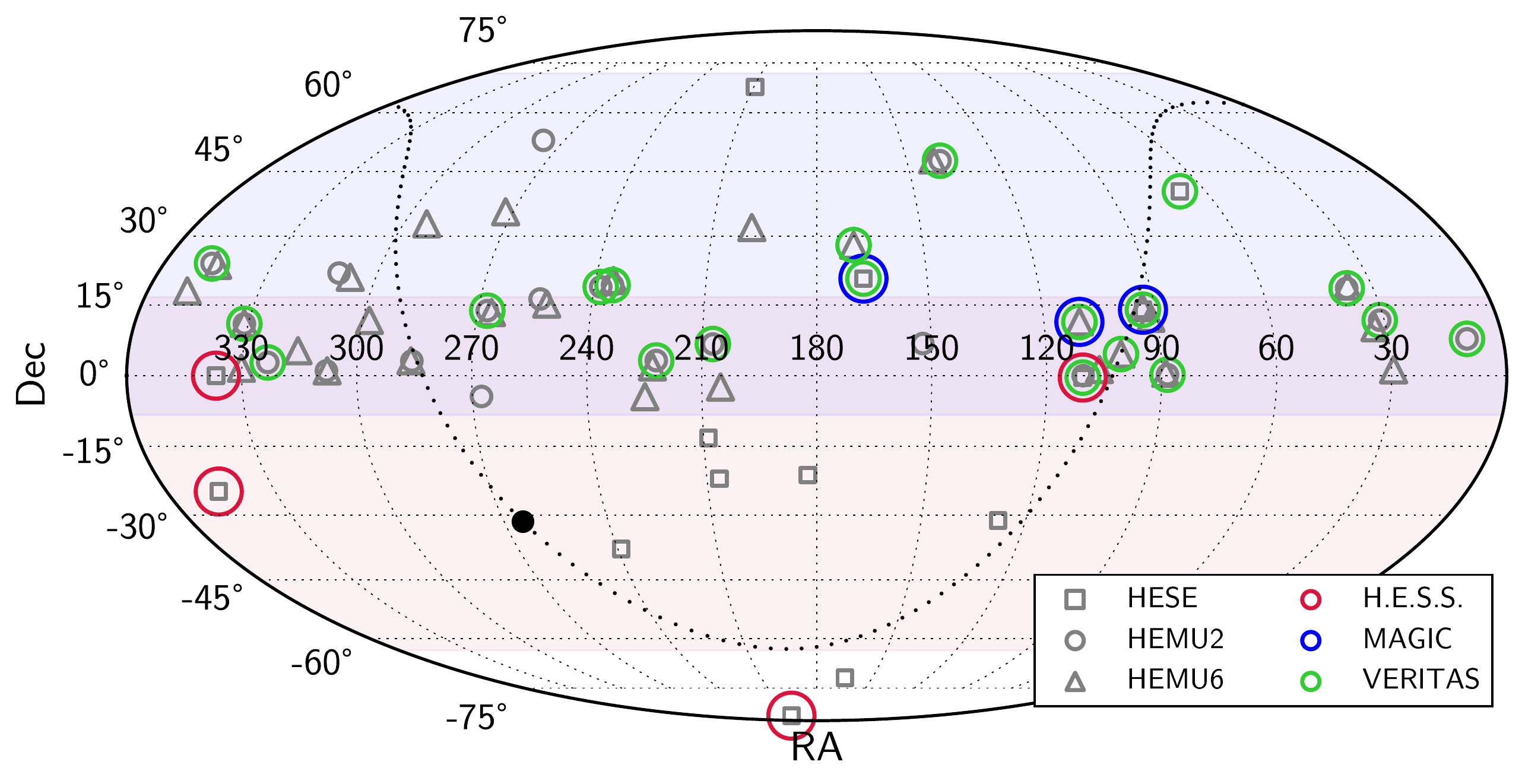}
 \caption{Sky map in equatorial coordinates of the neutrino event positions used in this program. In gray, different markers indicate the IceCube analyses from which the positions were obtained (HESE \cite{HESE3}, HEMU2 \cite{Chris}, and HEMU6, \cite{Leif}) while the colored markers indicate which IACT has performed follow-up observations of these events. The colored regions indicate the area of the sky observable at elevations $>50^{\circ}$ (typical for IACT observations) from the latitude of the FACT/MAGIC/VERITAS (blue) and H.E.S.S. (red) sites. The dotted black line represents the Galactic Plane, with a black circle indicating the Galactic Center.}
  \label{fig:eventmap}
\end{figure}

The follow-up observations of published event positions were performed using the H.E.S.S.~\cite{HESS}, MAGIC~\cite{MAGIC}, and VERITAS~\cite{VTS} arrays. A skymap of the neutrino positions indicating which IACT has performed follow-up observations is shown in Fig.~\ref{fig:eventmap}, and a list of the exposure times per IACT is given in Table~\ref{tab_events}. In April 2016, IceCube started a program that broadcasts a low-latency alert (typically below a minute) to follow-up partner observatories when a potentially-astrophysical neutrino event is detected at the South Pole and identified by an online selection algorithm~\cite{ICRealTime}. A description of the status of this program, which also involves the FACT project~\cite{FACT}, is given in Section~\ref{sec:discussion}.

The four observatories involved in this program use the imaging atmospheric Cherenkov technique, which uses large optical telescopes that collect the Cherenkov light produced by gamma-ray showers in the atmosphere and focuses it onto a sensitive cameras equipped with fast-readout light sensors (photomultiplier tubes in the case of H.E.S.S., MAGIC and VERITAS, and silicon photomultipliers in FACT). The geometry and brightness of the shower images is used to reconstruct the energy and incoming direction of the primary gamma-ray photon. IACT arrays use multiple telescopes to provide a stereoscopic view of the gamma-ray showers, which improves the angular resolution of the telescopes, typically of order $\sim0.1^{\circ}$ above 100 GeV. See~\cite{GammaReview} and references therein for a recent review of the IACT technique and the capabilities of current instruments.

\begin{table*}
\small
\centering
\begin{tabular}{ccccccc}
\hline
\textbf{ID} & \textbf{RA} & \textbf{Dec} & \textbf{Ang. res.} & \textbf{H.E.S.S. obs.} & \textbf{MAGIC obs.} & \textbf{VERITAS obs.} \\
 & \textbf{[$^{\circ}$]} & \textbf{[$^{\circ}$]} & \textbf{[$^{\circ}$]} & \textbf{time [min]} & \textbf{time [min]} & \textbf{time [min]} \\
\hline
HESE-05 & 	110.6 	& 	-0.4 		& 	$<1.2^{\circ}$ 		&		72		&		-		& 		205 	\\
HESE-13 & 	67.9 		& 	40.3		& 	$<1.2^{\circ}$ 		&		-		&		-		& 		242 	\\
HESE-18 & 	345.6 	& 	-24.8		& 	$<1.3^{\circ}$ 		&		486		&		-		& 		- 	\\
HESE-37 & 	167.3 	& 	20.7		& 	$<1.2^{\circ}$ 		&		-		&		447		& 		138 	\\
HESE-38 & 	93.3 		& 	14.0		& 	$<1.2^{\circ}$ 		&		-		&		342		& 		90 	\\
HESE-44 & 	336.7 	&	0.0		& 	$<1.2^{\circ}$ 		&		432		&		-		& 		- 	\\
HESE-45 & 	219.0 	&	-86.3		& 	$<1.2^{\circ}$ 		&		270		&		-		& 		- 	\\
\hline
HEMU2-02 & 	88.5 	&	0.2	& 	$\sim1^{\circ}$ 		&		-		&		-		& 		133 	\\ 
HEMU2-03 & 	37.1 	&	18.6	& 	$\sim1^{\circ}$ 		&		-		&		-		& 		180 	\\ 
HEMU2-05 & 	331.0 	&	11.0	& 	$\sim1^{\circ}$ 		&		-		&		-		& 		60 	\\ 
HEMU2-06 & 	346.8 	&	24.0	& 	$\sim1^{\circ}$ 		&		-		&		-		& 		115 	\\ 
HEMU2-07 & 	267.5 	&	13.8	& 	$\sim1^{\circ}$		&		-		&		-		& 		25 	\\ 
HEMU2-08 & 	238.3 	&	18.9	& 	$\sim1^{\circ}$ 		&		-		&		-		& 		30 	\\ 
HEMU2-09 & 	235.2 	&	19.3	& 	$\sim1^{\circ}$ 		&		-		&		-		& 		30 	\\ 
HEMU2-11 & 	323.3 	&	2.8	& 	$\sim1^{\circ}$		&		-		&		-		& 		40 	\\ 
HEMU2-13 & 	9.4 	&	7.8	& 	$\sim1^{\circ}$ 		&		-		&		-		& 		115 	\\ 
HEMU2-14 & 	207.2 	&	6.7	& 	$\sim1^{\circ}$		&		-		&		-		& 		76 	\\ 
HEMU2-19 & 	221.9 	&	3.2	& 	$\sim1^{\circ}$ 		&		-		&		-		& 		20 	\\ 
HEMU2-20 & 	138.9 	&	47.6	& 	$\sim1^{\circ}$		&		-		&		-		& 		60 	\\ 
HEMU2-21 & 	31.2 	&	11.8	& 	$\sim1^{\circ}$ 		&		-		&		-		& 		60 	\\ 
\hline
HEMU6-20 & 	169.61 	&	28.04	& 	$<0.5^{\circ}$ 		&		-		&		-		& 		50 	\\ 
HEMU6-27 & 	110.63 	&	11.42	& 	$<0.3^{\circ}$ 		&		-		&		648		& 		110 	\\ 
HEMU6-28 & 	100.48 	&	4.56	& 	$<0.3^{\circ}$ 		&		-		&		-		& 		80 	\\ 
\hline
\end{tabular} 
\caption{Neutrino track-like events observed as part of this program. The IDs represent the event numbers in the cited papers, while for HEMU2 it represents the position in the table available in \cite{ChrisTable}. The total observing time at each position after quality cuts for each IACT is shown in minutes.}
\label{tab_events}
\end{table*}

\section{Analysis and preliminary results}

The IACT observations summarized in Table~\ref{tab_events} were performed using the standard \emph{wobble} observation strategy where telescopes are offset from the position of the potential source to allow for a simultaneous determination of the background event rate. Offsets of $0.4^{\circ}-0.9^{\circ}$ with respect to the best-fit neutrino location were used to provide better coverage of the neutrino error circle. The analysis of the data was performed independently by each IACT group by introducing cuts to separate gamma-ray shower candidate events from the dominant background of hadronic cosmic-ray showers. The spatial distribution of events passing these cuts is studied to search for a point-like excess in the uncertainty region of each neutrino event that could represent evidence of VHE gamma-ray emission. The statistical significance of an excess is estimated using the method of Li \& Ma~\cite{LiMa} after events are smoothed using the IACT point spread function. 

\begin{figure*}[!th]
\centering
\includegraphics[width=0.7\textwidth]{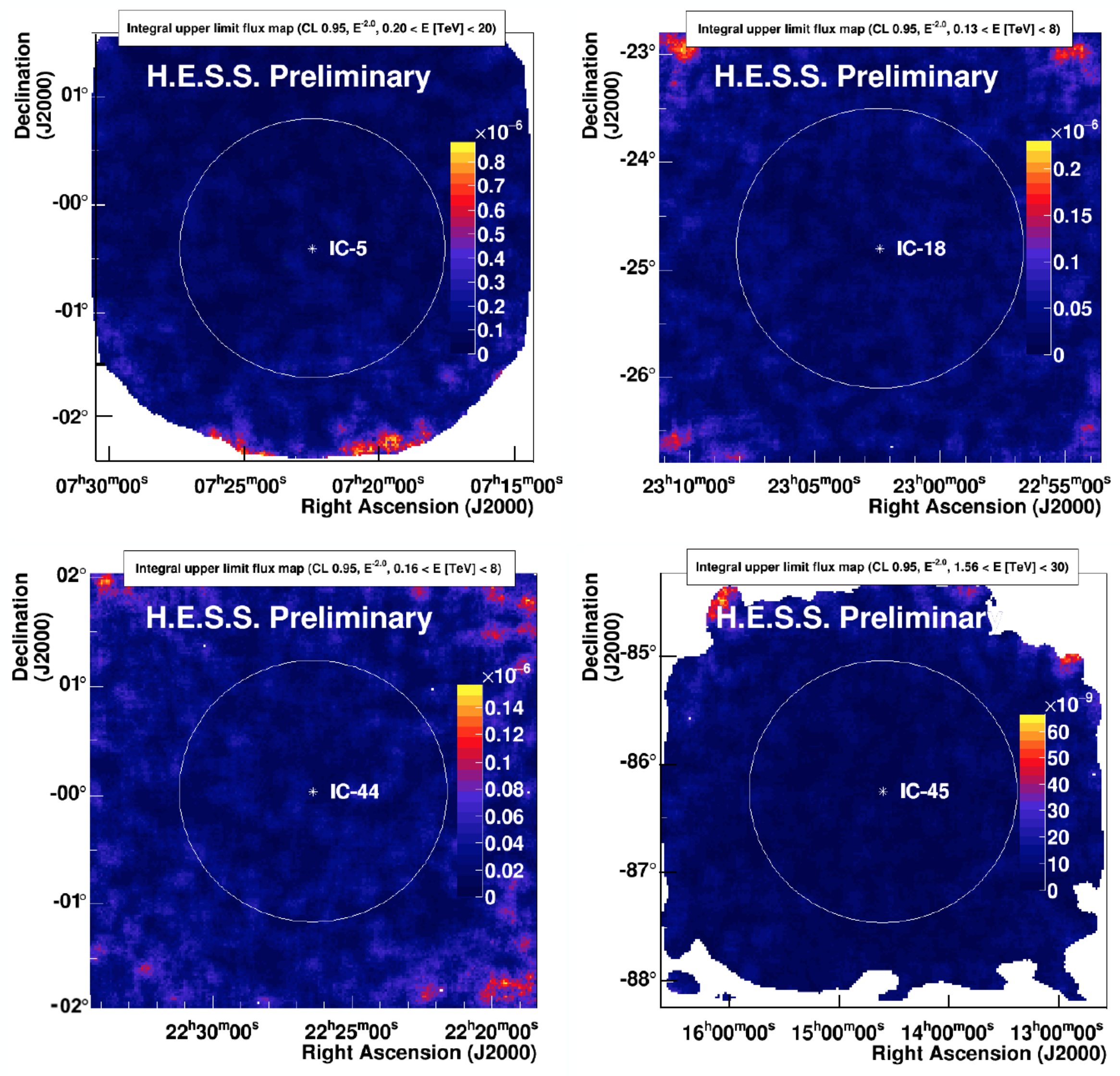}
 \caption{Maps showing upper limits on the gamma-ray flux derived from H.E.S.S. observations. The white circles denote the 50\% localisation uncertainty of the IceCube neutrino events.}
  \label{fig:HESSULs}
\end{figure*}

\begin{figure}[th]
\centering
\includegraphics[width=0.33\textwidth]{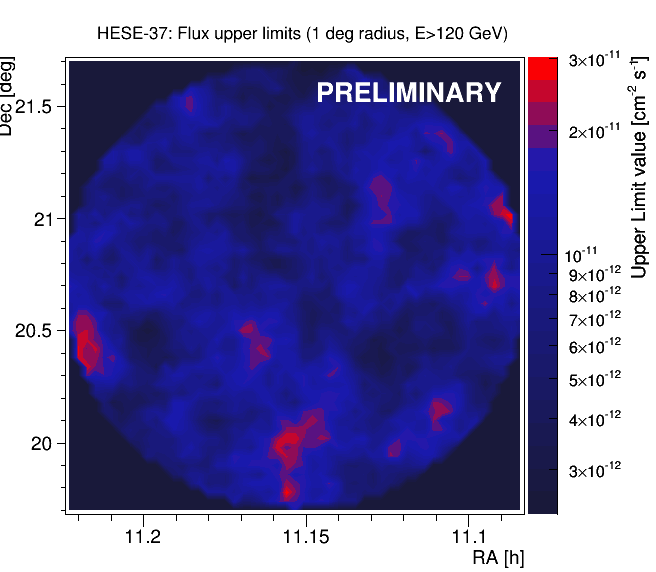}
\includegraphics[width=0.32\textwidth]{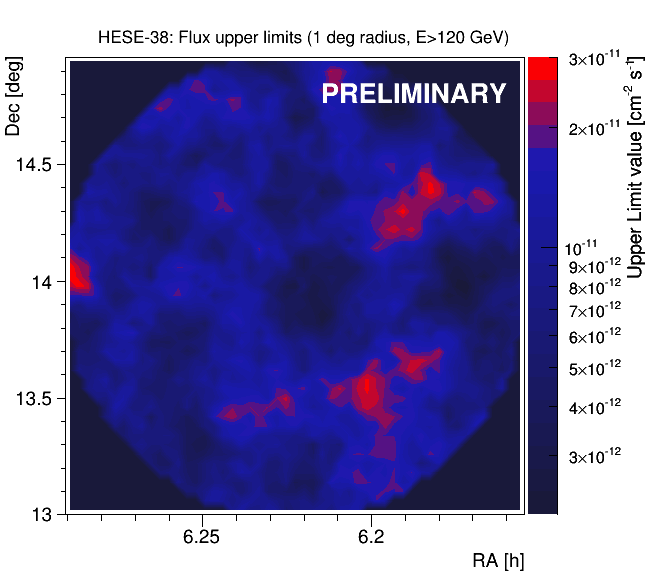}
\includegraphics[width=0.33\textwidth]{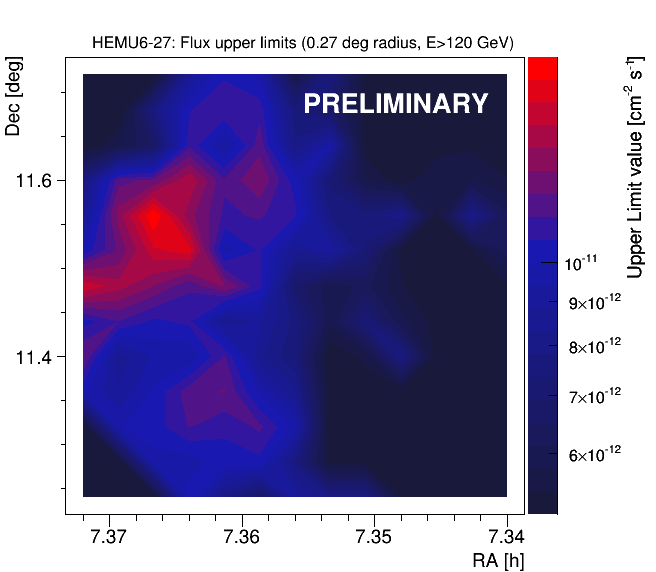}
 \caption{Sky maps showing integral flux upper limits for E$>$120 GeV derived from the MAGIC observations of HESE-37, HESE-38 and HEMU6-27. The size of the sky map corresponds to the IceCube 50\% error region of each event.}
  \label{fig_TSmap1}
\end{figure}

\begin{figure}[th]
\centering
\includegraphics[width=0.33\textwidth]{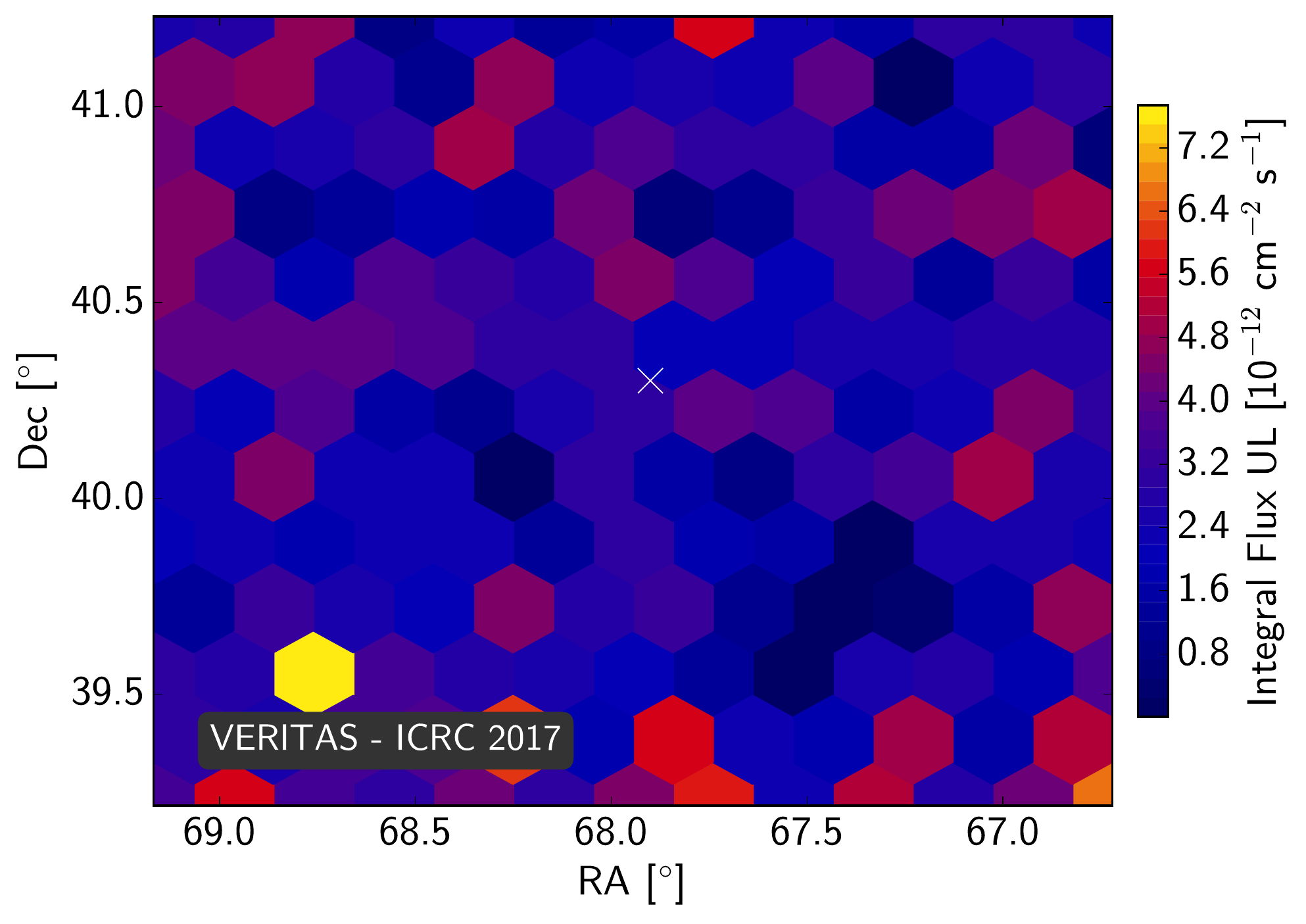}
\includegraphics[width=0.32\textwidth]{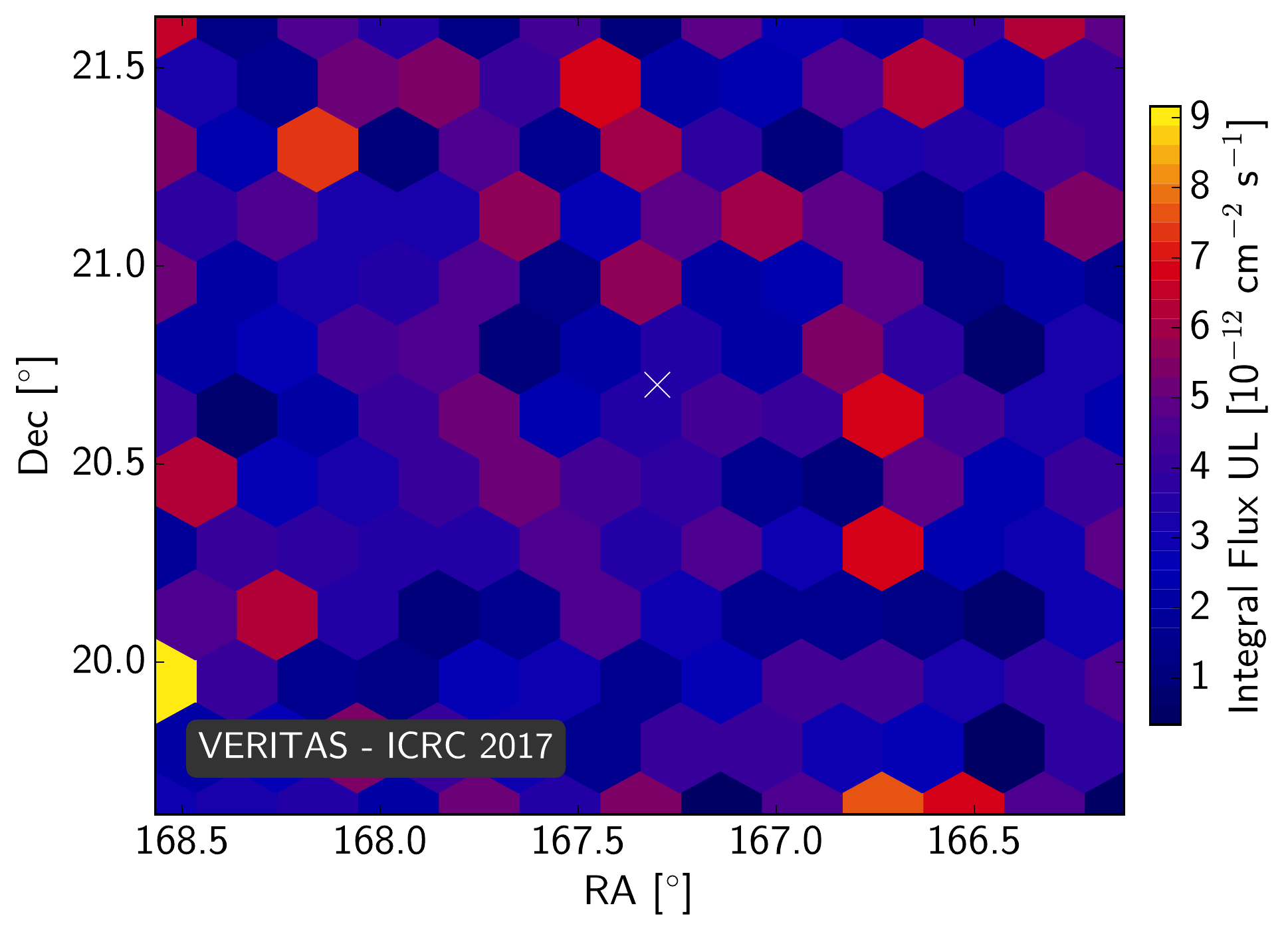}
\includegraphics[width=0.33\textwidth]{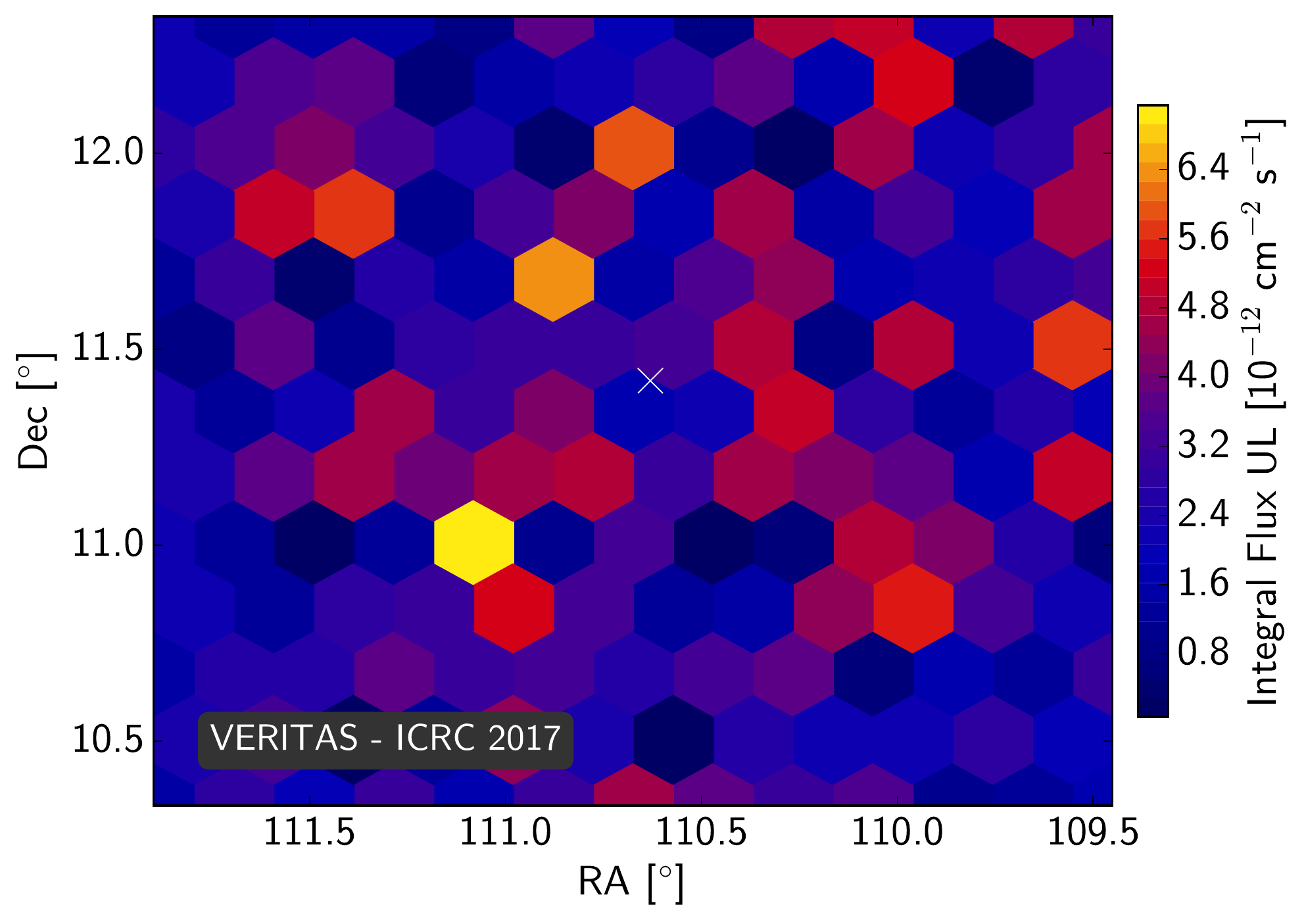}
 \caption{Sky maps showing VERITAS integral flux upper limits
 for HESE-13 (\emph{left}), HESE-37 (\emph{center}), and HEMU6-27 (\emph{right}) for E $>$ 170 GeV. The size of the sky map corresponds to the IceCube 50\% error region of each event.}
  \label{fig_ULmap_vts}
\end{figure}

No significant excess was found within any of the neutrino regions of interest, defined as the 50\% containment region of the neutrino events as given in Table~\ref{tab_events}, or anywhere in the IACT field of view. 
Given the large uncertainties of the neutrino event positions, integral upper limit maps above an energy threshold (which depends on the target) were also calculated covering the neutrino 50\% error circle to set constraints on potential gamma-ray sources. The energy threshold is calculated on a target-by-target basis as the energy at which the gamma-ray acceptance of the IACT array reaches 10\% of its maximum value. The upper bound of the validity range has been determined by the requirement of having more than 10 events in the OFF regions used to estimate the background. Upper limit maps at 95\% confidence level assuming a power-law spectrum following $E^{-2}$ for the potential sources were constructed. Examples for these flux upper limit maps for some of the targets are shown in Fig.\ref{fig:HESSULs} for H.E.S.S., Fig.~\ref{fig_TSmap1} for MAGIC, and Fig.~\ref{fig_ULmap_vts} for VERITAS.

\section{Discussion and Outlook}\label{sec:discussion}

The integral flux upper limits derived from IACT observations are considerably below the seven-year neutrino point-source sensitivity of IceCube, illustrating that even short IACT exposures can set strong constraints on the gamma-ray flux of any potential neutrino source detected by IceCube. 

The lack of VHE gamma-ray emission associated with the IceCube events can be interpreted as evidence of the gamma rays being absorbed before they reach Earth, either by the extragalactic background light (EBL) or due to interactions within the sources. The gamma-ray self absorption scenario has been postulated as the explanation for the possible tension between the \emph{Fermi} gamma-ray and IceCube neutrino diffuse backgrounds~\cite{Murase}. If the sources are optically-thin to gamma rays, the lack of correlation would indicate that the sources are too distant (absorbed) or too numerous (faint) to be detected in IACT follow-ups of individual neutrino events, at least for current instruments. The upper limit maps presented here serve two purposes: set constraints on any potential source in the vicinity of the neutrino event, and, in the case no significant source is found, enable a statistical search for systematic excesses associated with the neutrino pointings that are not present in observations of blank patches of sky. This will be the purpose of a future publication.

While the observation of these catalog events have only provided upper limits on steady-flux sources so far, theoretical models \cite{ProtonBlazar} suggest that the neutrino flux could correlate with the gamma-ray flux in the case of flaring events of active galactic nuclei, the most numerous type of VHE gamma-ray source. Therefore, it is also important to study the temporal correlation between neutrinos and gamma rays. This can be done either by correlating gamma-ray flaring events of  known sources with the time of neutrino detections~\cite{Kintscher}, or by trying to find electromagnetic counterparts of neutrino events in other wavebands via target-of-opportunity observations (e.g.~\cite{GFU}). For this, IceCube provides realtime alerts to the community using the gamma-ray coordinate network (GCN). Apart from the HESE events already described in this work, extremely-high-energy (EHE) events are also reported online. Observatories can subscribe to these alerts to perform immediate follow-up observations of the detected neutrino events. In that context, all IACTs discussed here started to perform follow-up observations of realtime alerts. First results are reported from FACT~\cite{FACTresults}, H.E.S.S.~\cite{HESS_MMalerts_ICRC17}, MAGIC~\cite{MAGIC_Gamma2016} and VERITAS~\cite{VERITAS_ICHEP2016}.

\section{Acknowledgements} 
\small

\noindent
Acknowledgements on behalf of the FACT Collaboration:\\ 
\noindent
\href{http://fact-project.org/collaboration/icrc2017_acknowledgements.html}{http://fact-project.org/collaboration/icrc2017\_acknowledgements.html}

\noindent
Acknowledgements for the H.E.S.S. Collaboration:\\
\noindent
\href{https://www.mpi-hd.mpg.de/hfm/HESS/pages/publications/auxiliary/HESS-Acknowledgements-ICRC2017.html}{https://www.mpi-hd.mpg.de/hfm/HESS/pages/publications/auxiliary/HESS-Acknowledgements-ICRC2017.html}

\noindent
Acknowledgements on behalf of the IceCube Collaboration:\\
\noindent
\href{http://icecube.wisc.edu/collaboration/authors/icrc17_icecube}{http://icecube.wisc.edu/collaboration/authors/icrc17\_icecube}

\noindent
Acknowledgements on behalf of the MAGIC Collaboration:\\
\noindent
\href{https://magic.mpp.mpg.de/acknowledgements_19_05_2017.html}{https://magic.mpp.mpg.de/acknowledgements\_19\_05\_2017.html}

\noindent
Acknowledgements on behalf of the VERITAS Collaboration: \href{https://veritas.sao.arizona.edu/}{https://veritas.sao.arizona.edu/}

\providecommand{\href}[2]{#2}\begingroup\raggedright\endgroup

\end{document}